\documentclass[aps,twocolumn,prl,tightenlines,floatfix,showpacs]{revtex4}
\usepackage{graphicx}
\usepackage[english]{babel}
\usepackage{amsmath}
\usepackage{amssymb}
\usepackage{times}

\begin{document}
\title{Controlling Transport of Ultra-Cold Atoms in 1D Optical Lattices with Artificial Gauge Fields}
\author{Chih-Chun Chien$^{1}$, and Massimiliano Di Ventra$^{2}$}

\affiliation{$^{1}$Theoretical Division, Los Alamos National Laboratory, MS B213, Los Alamos, NM 87545, USA \\
$^{2}$Department of Physics, University of California, San Diego, CA 92093, USA}
\date{\today}

\begin{abstract}
We show that the recently developed optical lattices with Peierls substitution -- which can be modeled as a lattice with a complex tunneling coefficient -- may be used to induce controllable quantum transport of ultra-cold atoms. In
particular, we show that by ramping up the phase of the complex tunneling coefficient in a spatially uniform fashion, a finite quasi steady-state current (QSSC) ensues from the exact dynamics of non-interacting fermions. The direction and magnitude of the current can be controlled by the overall phase difference but not the details of the ramp. The entanglement entropy does not increase when the QSSC lasts. Due to different spin statistics, condensed non-interacting bosons do not support a finite QSSC under the same setup. We also find that an approximate form of the QSSC survives when perturbative effects from interactions, weak harmonic background traps, and finite-temperature are present, which suggests that our findings should be observable with available experimental capabilities.
\end{abstract}

\pacs{05.60.Gg, 67.10.Jn, 72.10.-d}

\maketitle

There has been growing interest in studying quantum transport phenomena of ultra-cold atoms. For instance, recent experiments \cite{collision_tran,interaction_tran,slowmasstran,Blochtransport,Esslingertransport} have demonstrated how to induce a mass current using different confining potentials and drives, while theoretical studies \cite{Cramer08,MCFshort,int_induced,Bruderer12,Killi12} have explored interesting phenomena that may soon be tested experimentally. A mass current has been usually induced by a displacement or distortion of the confining potential \cite{collision_tran,interaction_tran,slowmasstran,Blochtransport}, a quench or imbalance of the density or interactions within the system \cite{Cramer08,MCFshort,int_induced,Killi12}, or connecting to external reservoirs \cite{Esslingertransport,Bruderer12}. At first glance it may then look counter-intuitive that one may induce a current in an isolated atomic cloud without either distorting the trap potential or by introducing any imbalance in its parameters. Here, instead we show the possibility of driving a current under those constraints by using the recently developed techniques of optical lattices with artificial (synthetic) gauge fields \cite{PS_exp,PS_exp2}.

Synthetic gauge fields for ultra-cold atoms have suggested interesting phenomena such as inducing vortices without rotating the atomic cloud \cite{Artificial_B}, oscillations of atomic clouds driven by synthetic electric fields \cite{Artificial_E}, and many others. Ref.~\cite{PS_exp} shows how a combination of an radio-frequency field and a Raman field can generate a lattice potential with the feature that atoms acquire a Berry phase when tunneling to a different site. This can be modeled as a lattice with the Peierls substitution \cite{Hofstadter76}, where the tunneling coefficient is generalized to a complex number $\bar{t}e^{i\phi}$. Ref.~\cite{PS_exp2} also presents optical lattices with complex tunneling coefficient by temporally modulating the lattice with designed patterns. As demonstrated in Refs.~\cite{PS_exp}, $\bar{t}$ and $\phi$ can be tuned separately in experiments. Here we generalize the concept of Refs.~\cite{PS_exp,PS_exp2} and use a micro-canonical formalism \cite{Bushong05,MCFshort} to study the atomic dynamics when $\phi$ is a function of time. We first consider non-interacting particles, an important but difficult situation to realize for electronic solid-state systems which can instead be easily accomplished in cold-atom systems \cite{ChinRMP}. This allows us to study the role of interactions in an unambiguous way. Then perturbative effects of interactions, finite temperatures, and weak harmonic background traps will be studied.

Our major findings are: {\it i)} By ramping up the phase, $\phi$, as a function of time $t$ in a \textit{spatially uniform} fashion, an atomic mass current can be induced in a closed isolated system with or without a weak harmonic trap potential. {\it ii)} The \textit{direction and magnitude} of the mass current can be controlled by the displacement of $\phi$. {\it iii)} For non-interacting fermions, a quasi-steady-state current (QSSC) and its approximate form persist for a period of time depending on the system size. A condensate of non-interacting bosons, in contrast, does not support a finite QSSC. {\it iv)} The fermionic QSSC exhibits no memory effects, which means that the magnitude of the QSSC is insensitive to the details of the ramp. {\it v)} The entanglement entropy remains constant when the QSSC lasts, and finally {\it vi)} The QSSC (and its approximate form) survives when perturbative effects from interactions and finite temperatures are considered. All these predictions can be tested with minimal modifications of available experimental techniques \cite{PS_exp,PS_exp2}.

Following Ref.~\cite{PS_exp}, we consider non-interacting ultra-cold atoms \cite{non_int} loaded into a finite one-dimensional optical lattice, which may be modeled by the Hamiltonian
\begin{equation}\label{eq:H}
H=-\sum_{j=1}^{N-1}[\bar{t}e^{i\phi}c^{\dagger}_{j}c_{j+1}+H.c.]
\end{equation}
with open boundary conditions. Here $\bar{t}$ and $\phi$ are the amplitude and the phase of the hopping coefficient and should be tunable separately \cite{PS_exp}, $c^{\dagger}_{j}$ ($c_j$) is the creation (annihilation) operator of site $j$, and $N$ is the lattice size. We assume $\phi$ and $\bar{t}$ are uniform across the whole lattice. The unit of time is chosen as $t_0\equiv \hbar/\bar{t}$.

The system can be driven out of equilibrium by varying $\phi$ over time. The current flowing from the left half $(j\le N/2)$ to the right half $(j>N/2)$ is given by $I=-dN_{L}/dt$, where $N_{L}=\sum_{j=1}^{N/2}\langle c^{\dagger}_{j}c_{j}\rangle$. Explicitly,
\begin{eqnarray}\label{eq:I}
I=2\mbox{Im}[\bar{t}e^{i\phi}\langle c^{\dagger}_{N/2}c_{N/2+1}\rangle].
\end{eqnarray}
The current can be measured by the protocol of Refs.~\cite{slowmasstran,MCFshort}.
The time evolution of the single-particle correlation matrix $c_{ij}(t)\equiv\langle c^{\dagger}_{i}(t)c_{j}(t)\rangle$ follows the equations of motion
\begin{eqnarray}\label{eq:EOM}
i\frac{\partial}{\partial t}c_{ij}(t)&=&\bar{t}e^{i\phi(t)}c_{i-1,j}(t)+\bar{t}e^{-i\phi(t)}c_{i+1,j}(t)- \nonumber \\
& &\bar{t}e^{i\phi(t)}c_{i,j+1}(t)-\bar{t}e^{-i\phi(t)}c_{i,j-1}(t).
\end{eqnarray}
For a closed system as the present one, $c_{ij}=0$ if $i,j < 1$ or $i,j > N$. We first consider $N_p$ non-interacting fermions loaded into the ground state of the lattice with $\phi=0$ initially.

\begin{figure}
  \includegraphics[width=3in,clip]
{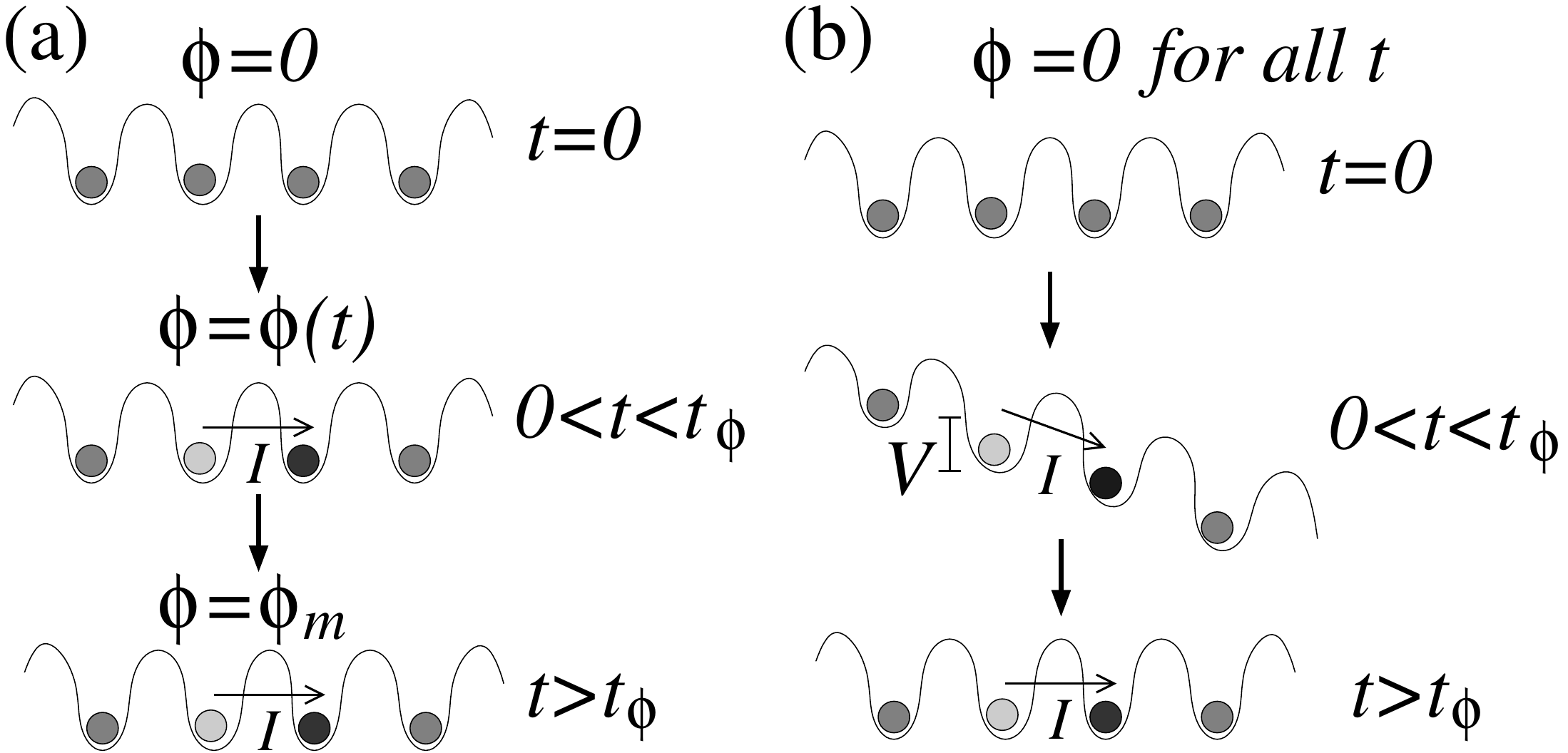}
  \caption{Schematics of the setups of fermions for (a) phase-induced transport and (b) an effective model for describing (a). Here grey dots imply that particles may be in a superposition of quantum states.}
\label{fig:PS_cartoon}
\end{figure}

We then let the phase $\phi(t)$ rise 
according to $\phi(t)=(t/t_{\phi})\phi_m$ when $t\le t_{\phi}$ and $\phi(t)=\phi_m$ when $t>t_{\phi}$. Note that $\phi(t)$ is spatially uniform across the whole lattice at each time $t$. Figure~\ref{fig:PS_cartoon} (a) illustrates the setup.
The unitary transformation $c_{j}=\sum_{k=1}^{N}U_{jk}b_{k}$ with $U_{jk}=\sqrt{\frac{2}{N+1}}\sin\left(\frac{jk\pi}{N+1}\right)e^{i(j-1)\phi}$ leads to $H=\sum_{k=1}^{N}E_{k}b_{k}^{\dagger}b_{k}$ with $E_{k}=-2\bar{t}\cos\left(\frac{k\pi}{N+1} \right)$. We remark that $U_{jk}$ depends explicitly on $\phi$ while $E_{k}$ is independent of $\phi$. In the energy basis, the initial state of fermions corresponds to $\langle b^{\dagger}_{k}b_{k^{\prime}}\rangle=\delta_{kk^{\prime}}\theta(N_{p}-k)$ with $\phi=0$, where $\theta(x)=1$ if $x\ge 0$ and $\theta(x)=0$ otherwise. The corresponding $c_{ij}(t=0)$ can be inferred from the unitary transformation.

In addition to the current, we also evaluate the entanglement entropy, $s$, between the left half and the right half. The entanglement entropy for a bipartite system with noninteracting fermions has been discussed in Ref.~\cite{KlichLevitov}. Here it can be calculated as follows. We define $M_{ij}(t)=c_{ij}(t)$ for $1\le i,j \le (N/2)$ as the sub-matrix of the correlation matrix for the left half with the eigenvalues $v_{j}(t)$. Then 
\begin{eqnarray}\label{eq:s}
s(t)=-\sum_{j=1}^{N/2}[v_{j}(t)\log v_{j}(t)+(1-v_{j}(t))\log(1-v_j(t))].
\end{eqnarray}
For electrons modeled as noninteracting particles flowing through a quantum point contact \cite{KlichLevitov}, the rate at which $s(t)$ increases depends on the transmission coefficient, which measures the probability that a particle can tunnel from the left half to the right half when the systems exhibit a steady or quasi-steady current.

\begin{figure}
  \includegraphics[width=3in,clip]
{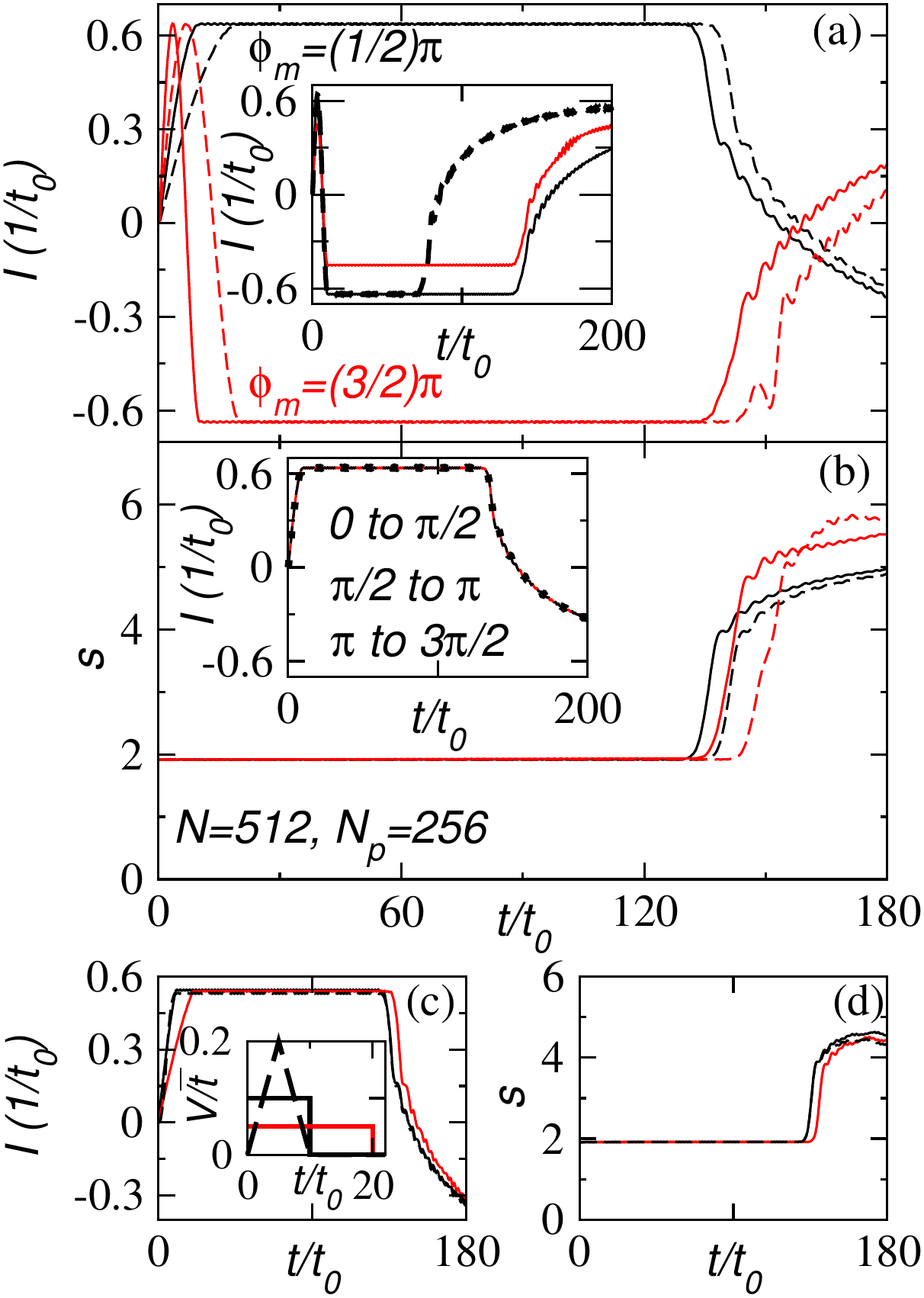}
  \caption{(Color online) The fermionic current (a) and entanglement entropy (b) from phase-induced transport. We show $\phi_m=\pi/2$ in black and $\phi_m=(3/2)\pi$ in red. Solid and dashed lines correspond to $t_{\phi}=10t_0$ and $20t_0$, respectively. Here $N=512$ and $N_p=256$. The inset of (a) shows how the quasi steady-state current varies as $N$ or $N_p$ changes. The black solid line, red solid line, and black dashed line correspond to $(N,N_p)=(512,256)$, $(512,128)$, and $(256,128)$ with $\phi_{m}=(3/2)\pi$ and $t_{\phi}=10$. The inset of (b) shows the currents by ramping up $\phi$ from $0$ to $\pi$ (black dashed line), from $\pi/2$ to $\pi$ (red solid line), and from $\pi$ to $3\pi/2$ (black dotted line) for $N=512$ and $N_p=256$ with $t_{\phi}=10t_0$. All three curves coincide. The current and entanglement entropy of the effective model \eqref{eq:H} with $N=512$, $N_p=256$, and $t_{\phi}=10$ are shown in (c) and (d). The black solid lines, red solid lines, and black dashed lines correspond to the three selected $V_0(t)$ shown in the inset of (c). Note that $\int_{0}^{t_\phi}V_0(t)dt$ are the same for all three cases.}
\label{fig:Iands}
\end{figure}

The current and entanglement entropy of a system with $N=512$ and $N_p=256$ are shown in Figure~\ref{fig:Iands} for $\phi_{m}=\pi/2$ and $(3/2)\pi$ with $t_{\phi}=10t_0$ and $20t_0$. We consider a finite ratio of $N_p/N$ as discussed in Ref.~\cite{TDlimit}. Due to the finite size of the system, the wavefunctions will interfere with themselves after some revival time, as one can see on Fig.~\ref{fig:Iands}. Throughout the paper we will focus on the physics before this revival time, which we found to be linearly proportional to the system size.

A key point is that even when the lattice, initial density distribution, and $\phi(t)$ are all spatially uniform, there can be a current flowing across the middle of the lattice. The mechanism behind this current is because \textit{the initial ground state for $\phi=0$ is not the ground state for $\phi > 0$} so the spreading of the wavefunction leads to a mass current. One can clearly see that for the currents there is a plateau for each curve in a given time interval. This plateau resembles the quasi-steady-state current (QSSC) -- studied in Refs.~\cite{Bushong05,MCFshort} -- induced by an inhomogeneous bias or density. We found that for a fixed ratio of $N_p/N$, the duration of the QSSC scales linearly with $N$ and its magnitude is the same. For a fixed $N$, the duration of the QSSC is the same as $N_p$ varies but the magnitude decreases in a non-linear fashion as $N_p$ decreases. The inset of Fig.~\ref{fig:Iands} (a) shows the currents for $(N,N_p)=(512,256)$, $(512,128)$, and $(256,128)$ with $\phi_{m}=(3/2)\pi$ and $t_{\phi}=10t_0$ and demonstrates those features \cite{pBC_note}. 

Importantly, \textit{the direction and magnitude} of the QSSC can be controlled by $\phi_m$ for fixed $N$ and $N_p$. We also remark that the QSSC lasts when $d\phi/dt=0$ until the revival time and this is partially due to the fact that there is no dissipation in an isolated non-interacting system. Another important feature is that the magnitude of the QSSC for fixed $N$ and $N_p$ is insensitive to different ways of turning on the phase $\phi(t)$ \cite{phi_note}. Our findings suggest that the system exhibits no memory effect when the QSSC has been developed and maintained.

The entanglement entropy remains the same when the QSSC lasts, which is consistent with previous studies with different setups \cite{KlichLevitov} in the limit where the transmission coefficient is equal to one due to the uniform tunneling coefficient. One may understand this behavior by the full counting statistics which models the particles semi-classically with probability $\mathcal{T}$ (equal to the transmission coefficient) of tunneling through the middle and with probability $1-\mathcal{T}$ of being reflected \cite{KlichLevitov}. The entanglement entropy increases because this probability distribution introduces correlations between the right half (with tunneling particles) and the left-half (with reflected particles). For a uniform lattice without any scattering potential, however, the transmission coefficient is $\mathcal{T}=1$ \cite{Maxbook}, so particles tunnel through the middle without building any correlations between the two halves of the lattice.

Let us now analyze the effect of ramping up $\phi$ from different initial values. The corresponding initial states are determined by diagonalizing the Hamiltonian \eqref{eq:H} with different $\phi(t=0)$ and constructing the ground states. The currents for different initial and final values of $\phi(t)$ are shown in the inset of Fig.~\ref{fig:Iands} (b). We deliberately choose $(\phi(t=0),\phi_m)$ as $(0, \pi/2)$, $(\pi/2,\pi)$, and $(\pi, 3\pi/2)$ with the same ramp time $t_{\phi}=10t_0$. All three curves coincide. This can be explained by the fact that in quantum mechanics, an overall constant phase change of the wavefunction does not introduce observable effects. Therefore the dynamics only depends on the displacement of the phase. One may notice that the QSSC from $\phi(t=0)=0$ to $\phi_m=3\pi/2$ has the opposite sign of that from $0$ to $\pi/2$. To observe this reversal of the direction of the QSSC, one has to ramp the phase by the full amount. Simply going from $\pi$ to $3\pi/2$ does not show this reversed current. The ability to reverse the direction of the current is because the factor $e^{i\phi}$ in the Hamiltonian \eqref{eq:H} selects the preferred direction of tunneling. Since $e^{i(\phi+\pi)}=-e^{i\phi}$, the preferred direction can be reversed. One may also see this reversion of the current from the expression of $I$ shown in Eq.~\eqref{eq:I}. This relation also implies that there is no current when $\phi_m=\phi(t=0)+\pi$, which we have verified numerically.

To understand why varying $\phi(t)$ can drive a QSSC, we resort to a classical analogy. Using $\mathbf{E}=-\partial \mathbf{A}/\partial t$ and $I=V/R$, one expects that the instantaneous current is linearly proportional to $d\phi/dt$, where $\phi=\int\mathbf{A}\cdot d\mathbf{l}$. Here $\mathbf{E}$ is an electric field causing the voltage difference $V$ and $R$ is the resistance of the classical system.
We thus assume that $d\phi/dt$ induces a voltage difference $V$ between adjacent sites $j$ and $j+1$ and model the quantum system by the effective Hamiltonian
\begin{eqnarray}\label{eq:Heff}
H_{eff}=-\bar{t}\sum_{j=1}^{N-1}(c^{\dagger}_{j}c_{j+1}+c^{\dagger}_{j+1}c_{j})+\sum_{j=1}^{N-1}V_j(t)c^{\dagger}_{j}c_{j}.
\end{eqnarray}
Here $V_j(t)=0$ if $t\le 0$, $V_j(t)=V_0(t)(N-j)$ during $0<t\le t_{\phi}$, and then $V_j(t)=0$ for $t>t_{\phi}$. In other words, the lattice is tilted during $0<t\le t_{\phi}$ with a potential difference $V_0(t)$ between adjacent sites. In the effective model (\ref{eq:Heff}) the tunneling coefficient is real and time-independent. Figure~\ref{fig:PS_cartoon} (b) illustrates this effective model.

Figure~\ref{fig:Iands} (c) and (d) show the current and entanglement entropy from the effective Hamiltonian \eqref{eq:Heff} for several functional forms of $V_0(t)$ shown in the inset. Two important features immediately connect the effective model and the phase-induced transport: {\it i)} a QSSC and {\it ii)} a constant entanglement entropy when the QSSC persists. Therefore it is the potential difference due to $d\phi/dt$ during $t_{\phi}$ that drives the QSSC and the QSSC can continue after $d\phi/dt=0$ or the potential difference vanishes. Another important feature is that for different functional forms of $V_0(t)$ with the same $\int_{0}^{t_{\phi}}V_0(t)dt$, the magnitude of the QSSCs are extremely close to each other. This again exhibits the lack of memory effects in the transport 
of non-interacting systems because the QSSC is controlled by the area enclosed by $V_0(t)$, not the detailed form of $V_0(t)$. However, our results bear a stark difference between the classical and quantum systems: For the quantum system, the QSSC is controlled by the total difference of $\phi$, not its time derivative.

In Ref.~\cite{MCFshort} it was shown that condensed noninteracting bosons cannot support a QSSC when the system is subject to a sudden density imbalance. For bosons the Hamiltonian and the equations of motion are identical to those of fermions. At zero temperature, all non-interacting bosons occupy the lowest energy state so the initial condition is $\langle b^{\dagger}_{k}b_{k^{\prime}}\rangle=N_p \delta_{kk^{\prime}}\delta_{k,1}$ with $\phi=0$, where $k=1$ denotes the lowest energy state. We did not find any plateau in the bosonic current, namely initially condensed non-interacting bosons do not support a QSSC although a time-dependent $\phi(t)$ can still drive a finite current \cite{boson_note}.

\begin{figure}
  \includegraphics[width=3.4in,clip]
{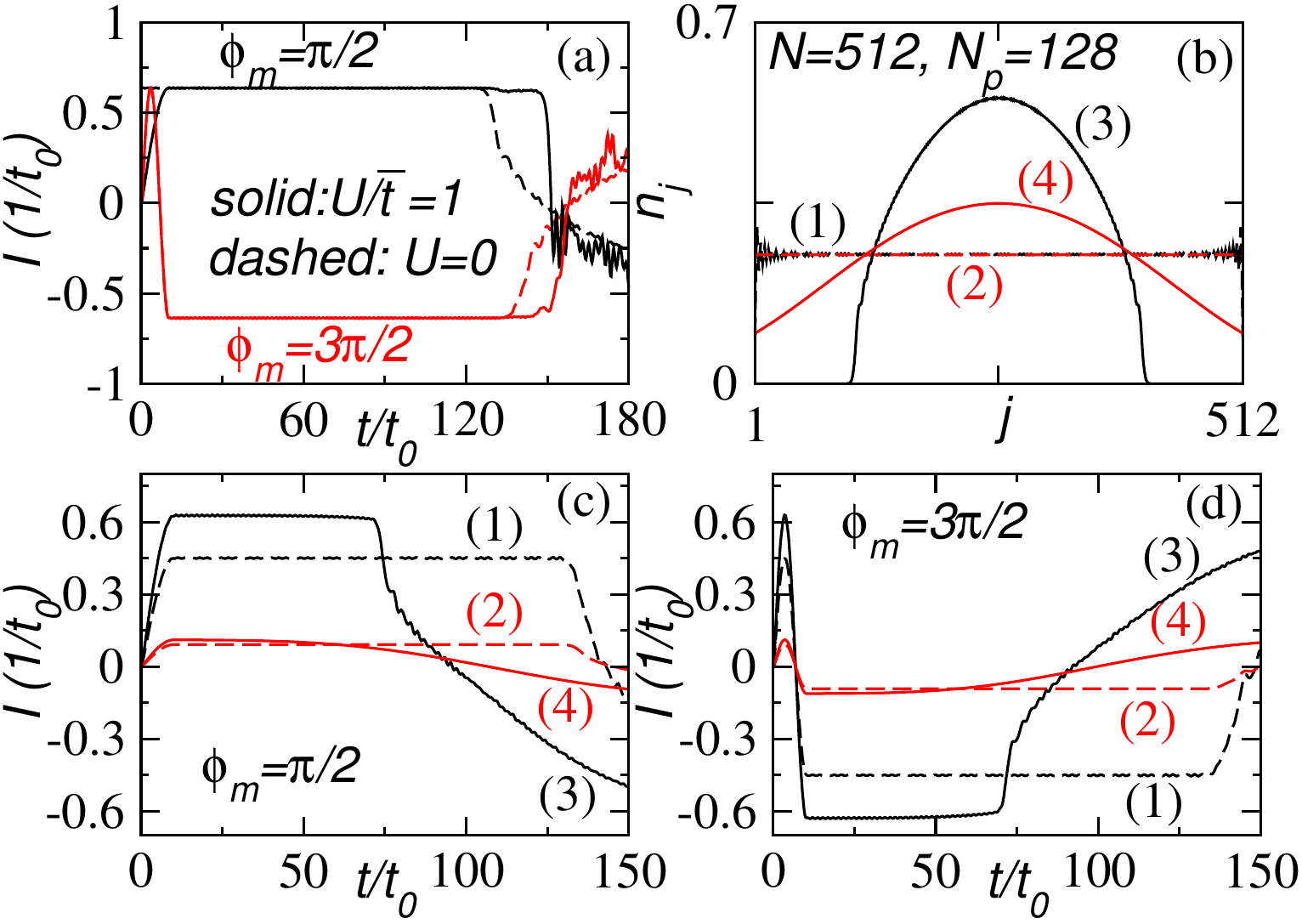}
  \caption{(Color online) The phase-induced currents under different conditions: (a) Effects of interactions. We show the currents for one species of a two-component Fermi gas with onsite interactions $U/\bar{t}=1$ (solid lines) and without interactions (dashed lines).  Here $\phi_m=\pi/2$ (black) and $\phi_m=3\pi/2$ (red) with $t_{\phi}=10t_0$, $N=512$, and $N_p=256$. (b) The initial ground-state density profiles for different confining potentials and temperatures: (1) and (2) correspond to the system without a harmonic potential, while (3) and (4) correspond to the system with a harmonic potential with $y=10^{-4}$. The temperatures of (1) and (3) are $T=0$ while those for (2) and (4) are $T=4\bar{t}/k_{B}$.
(c) and (d) show the currents from the four cases shown in (b). The maximal phase in (c) is $\phi_m=\pi/2$ and that in (d) is $3\pi/2$. The ramp time is $t_{\phi}=10t_0$ for both (c) and (d).   Here $N=512$ and $N_p=128$ for (b), (c), and (d).}
\label{fig:robust}
\end{figure}

Finally we consider other effects that may arise in actual experiments. The first case involves two-component fermions with weak onsite repulsive interactions of the form $H_{U}=U\sum_{j}n_{j\sigma}n_{j\bar{\sigma}}$, where $n_{j\sigma}=c^{\dagger}_{j\sigma}c_{j\sigma}$, $\sigma=1,2$ for the two species, and $\bar{\sigma}$ is the opposite of $\sigma$. The equations of motion with the standard Hartree-Fock approximation have been reported in Ref.~\cite{TDlimit}. We consider an initial state of the approximate Hamiltonian for the species $\sigma$: $H_{a}=-\bar{t}\sum_{j=1}^{N-1}(c^{\dagger}_{j}c_{j+1}+H.c.)+U\langle n_{\bar{\sigma}}\rangle n_{j\sigma}$, where $\langle n_{\bar{\sigma}}\rangle=(1/N)\sum_{j}\langle n_{j\bar{\sigma}}\rangle$. The unitary transformation $c_{j\sigma}=\sum_{k}\tilde{U}_{jk}\tilde{b}_{k\sigma}$ that diagonalizes $H_{a}$ can be found and the initial state corresponds to $\langle \tilde{b}^{\dagger}_{k\sigma}\tilde{b}_{k^{\prime}\sigma^{\prime}}\rangle =\delta_{kk^{\prime}}\delta_{\sigma\sigma^{\prime}}\theta(N_p-k)$, where we assume that there are $N_p$ particles for each species. Then the system evolves according to the Hamiltonian $H+H_{U}$ with a linear rising $\phi(t)$ for $t\le t_{\phi}$ and $\phi=\phi_m$ for $t>t_{\phi}$. We also assume that $\langle c^{\dagger}_{j\sigma}c_{j\sigma}\rangle = \langle c^{\dagger}_{j\bar{\sigma}}c_{j\bar{\sigma}}\rangle$ for all $t$. Figure~\ref{fig:robust} (a) shows the currents for $\sigma=1$ from the case with $U/\bar{t}=1$ and from the non-interacting case with $N=512$ and $N_p=256$. One can see that for weak interactions where the mean-field (Hartree-Fock) approximation is reasonable, the current is not affected qualitatively by the presence of the repulsive interactions.
%This is different from the case where atoms flow into an initially blocked vacuum region studied in Ref.~\cite{TDlimit},
%where repulsive interactions reduce the magnitude of the QSSC.
We remark that  the initial state is not the ground state of the full mean-field Hamiltonian $H+H_U$. Therefore even if $\phi(t)=0$ for all $t$, there is still a tiny current $(\sim 10^{-9}t_0^{-1})$ induced by the difference between the Hamiltonians. This is, however, a negligible effect.

Other two possible effects involve the background harmonic trap for holding the atomic cloud and finite temperature. We study them for the non-interacting single-species fermions case. The harmonic trap potential introduces a term $\sum_{j}(1/2)m\omega^2 d^2(j-N/2-1/2)^2 c^{\dagger}_{j}c_{j}$ to the Hamiltonian, where $m$ is the mass of atoms, $\omega$ is the trap frequency, and $d$ is the lattice constant. We let $y\equiv (1/2)m\omega^2 d^2/\bar{t}$. In the presence of a harmonic trap, the density is higher at the minimum of the trap potential and is lower towards the trap edge. If the density reaches $n_j=1$ at the center ($j$ around $N/2$), that region will become a band insulator and no current can flow through it in a one-band model. Therefore $N_p/N$ and $y$ cannot be too large or the insulating state at the center will emerge and block any current. We found that for a lattice with $N=512$ and $N_p=128$ held by a harmonic trap potential with $y=1\times 10^{-4}$, the maximal density at the center is less than one. Fig.~\ref{fig:robust} (b) shows the density profiles with and without a harmonic trap potential. The QSSC can still be observed at $T=0$ when a weak harmonic trap potential is present, albeit it lasts for a shorter time period compared to the case without the trap potential as shown in Fig.~\ref{fig:robust} (c) and (d).

To study finite-temperature effects, we first diagonalize the $\phi=0$ Hamiltonian with or without the trapping potential to obtain $H_{tot}=\sum_{k=1}^{N}E^{t}_{k}(b^{t})^{\dagger}_{k}b^{t}_{k}$. The initial condition is chosen as $\langle (b^{t})^{\dagger}_{k}b^{t}_{k^{\prime}}\rangle=f(E^{t}_{k})\delta_{kk^{\prime}}$, where $f(x)=(\exp[(x-\mu)/k_B T]+1)^{-1}$ is the Fermi distribution function. The chemical potential is determined by $N_p=\sum_{k=1}^{N}f(E^{t}_{k})$.  One can use the unitary transformation that diagonalizes the Hamiltonian to obtain $c_{ij}(t=0)$ and follow the equations of motion as $\phi(t)$ turns on. The currents from the initial states at $T=4\bar{t}/k_{B}$ with and without a harmonic trap potential are shown in Fig.~\ref{fig:robust} (c) and (d) for $\phi_{m}=\pi/2$ and $3\pi/2$ with $t_{\phi}=10t_0$. One can see that the phase-induced current at finite $T$ is smaller when compared to the current at $T=0$. This implies that the phase-induced current is a phenomenon that requires coherence of the initial state. At finite $T$ the initial state is mixed according to the Fermi-Dirac distribution and the tuning of the tunneling coefficient could not drive a current efficiently. Moreover, in the presence of a harmonic trap potential at finite $T$, we observe a slightly tilted regime in the current instead of a plateau that defines a QSSC. Therefore in experiments with a harmonic trap, there should only be an approximate QSSC where the current changes relatively slowly as time evolves and this approximate QSSC represents the QSSC that is more prominent as $T\rightarrow 0$. Nevertheless, for zero or finite temperatures with or without a harmonic potential, the direction and magnitude of the phase-induced current could still be controlled by the maximal phase $\phi_m$ even when $\phi(t)$ and $\bar{t}$ are uniform across the whole lattice. 

We have thus shown that controllable phase-induced transport should be feasible in optical lattices with Peierls substitutions in realistic experimental conditions. Combined with reservoirs \cite{Esslingertransport}, density- or interaction-imbalance \cite{MCFshort,int_induced}, and other types of proposed devices \cite{atomtronics}, phase-induced transport may find applications in atomtronics \cite{atomtronics} or quantum quench dynamics \cite{PolkovnikovRMP} and provides an interesting example of dynamical quantum phenomena.

CCC acknowledges the support of the U. S. DOE through the LANL/LDRD Program.
MD acknowledges support from the DOE grant DE-FG02-05ER46204 and UC Laboratories.

\bibliographystyle{apsrev}
%\bibliography{reference}

\begin{thebibliography}{26}
\bibitem[{non()}]{non_int}
\bibinfo{note}{Spin-polarized fermionic atoms are virtually non-interacting due
  to Pauli exclusion principle and $s$-wave scattering.}
\bibitem[{pBC()}]{pBC_note}
\bibinfo{note}{For periodic boundary conditions we found that varying $\phi$ as
  a function of time can induce an oscillating circulating current instead of a
  QSSC.}
\bibitem[{phi()}]{phi_note}
\bibinfo{note}{Here we show the results from linear ramps of $\phi$. We have
  checked more complicated ways of ramping $\phi$ and found the same
  conclusions.}
\bibitem[{bos()}]{boson_note}
\bibinfo{note}{This is consistent with Ref.~\cite{PS_exp2}, where simultaneous
  ramps of $\bar{t}$ and $\phi$ together result in oscillations.}
\end{thebibliography}

\end{document}